\documentstyle[12pt,epsfig]{article}
\date{}
\title{ $\gamma$ - beam propagation in the  anisotropic medium } 

\author{V.A.Maisheev \thanks{E-mail maisheev@mx.ihep.su} \\
{\it Institute for High Energy Physics, 142284, Protvino, Russia }}

\begin{document}
\maketitle
\def\arcctg{\mathop{\rm arcctg}\nolimits} 
\def\ch{\mathop{\rm ch}\nolimits}
\def\sh{\mathop{\rm sh}\nolimits} 
\def\Im{\mathop{\rm Im}\nolimits}
\def\Re{\mathop{\rm Re}\nolimits}

\begin{abstract}
 Propagation of $\gamma$-beam in the anisotropic medium is considered. 
The simplest example of such a medium of the general type is a combination 
of the two linearly polarized monochromatic laser waves with different 
frequencies (dichromatic wave). The optical properties of this combination 
are  described  with  the  use  of  the  permittivity  tensor.  The 
refractive indices 
and polarization characteristics of normal electromagnetic waves propagating in the
anisotropic medium  are found. The relations, describing variations of 
$\gamma$ -beam intensity and Stokes parameters as functions of the propagation
length, are obtained. The influence of laser wave intensity on the
propagation process is calculated. 

 The $\gamma$-beam intensity losses  in the dichromatic wave depend on the
initial circular polarization of  $\gamma$-quanta. This effect is also 
applied to single crystals which are oriented in some regions of coherent
pair production. In principle, the single crystal sensitivity to
a circular polarization can be used for determination of polarization of
high energy ( in tens GeV and more) $\gamma$-quanta and electrons. 
\end{abstract}

\section{ Introduction}
Polarization effects \cite{LL_ME,AG} arising from the  
visible light propagation 
in the anisotropic or gyrotropic medium are well-known. 
The theory \cite{C} makes prediction about the analogous effects for
$\gamma$-quanta with energy $ > 1$ GeV propagating in  single crystals,
which present the anisotropic medium. The main absorption process of 
$\gamma$-quanta in the single crystals is the electron-positron
pair production. The cross section of this process depends on  
the linear
polarization of $\gamma$-beam with respect to crystallographic planes.
As a result,  the monochromatic linearly polarized $\gamma$-beam may be
presented as a superposition of two electromagnetic waves with 
different refractive indices due to which the transformation of linear 
(circular) polarization to circular (linear) one  takes place.
 
 On the other hand, the process of $e^+ e^-$-pair production in single
crystals  is similar to the same  process  in  a  linearly  polarized 
laser wave \cite{BKS}. A possibility to use a bunch of linearly polarized
laser photons as a "single crystal" is pointed in \cite{MMF}, but
the concrete calculations of this possibility are not given.

  In the recent paper \cite{KS} it has been shown that the polarization effects
such as birefringence and rotation of polarization plane for $\gamma$- beams
with energies of tens GeV or more take  place for  short (about some
picoseconds) laser bunches and parameters of lasers, which may be provided
by real techniques. In the cited paper the differential  equations  which 
determine the variation of Stokes parameters and intensity of
$\gamma$-quanta traversing a bunch of arbitrary polarized laser photons
are obtained.
For these calculations  the well-known scattering amplitudes for the process of elastic 
scattering light by light \cite{B_DE,LL_KE} were used.
  Then in paper \cite{MV} the process of $\gamma$-beam  propagation in
the field of a laser wave was  investigated with the use of traditional
optical methods. Bisides, the transformation of $\gamma$-beam linear
polarization into circular in the anisotropic medium was discussed in \cite{AS}.

 The case of $\gamma$-quanta propagation in single crystals described in
\cite{C} is a special case of such a process. The general case of 
$\gamma$-propagation in single crystals oriented in the angle 
region corresponding
to process of the coherent pair production was considered in Ref.(\cite{MMF}). 
In this paper it was shown that the  propagating $\gamma$-beam  is a 
superposition of the two elliptically polarized waves but the description
of variation of the $\gamma$-beam polarization is unavailaible.

 In the present paper we study the general case of 
  high energy $\gamma$-quanta propagation through the anisotropic medium.
The anisotropic medium is determined as such a  medium, which optical
properties may be described using a symmetric permittivity 
tensor \cite{LL_ME,AG}. As will be indicated the simplest example of the
anisotropic medium of the general type is a superposition of the two
linearly polarized laser waves with different frequencies moving in
the same direction. We will study in detail the $\gamma$-quanta propagation in
such a combined laser wave with the goal of better understanding of this
process in more complicated cases, as in single crystals.

\section{ Permittivity tensor in anisotropic medium}

 We write the equations of the electromagnetic field in a medium
($\gamma$-beam propagating in a laser wave, single crystal, and the like)
in the following form \cite{LL_ME,AG}
\begin{eqnarray}
rot \vec{B} = {{1}\over {c}} {{\partial \vec{D}}\over {\partial t}}
 \, ,\; \; \; div \vec D =0 \, , 
  \nonumber \\
 rot \vec E = - {{1}\over {c}} {{\partial \vec B } \over {\partial t}} 
\, , \; \; \; div \vec B =0 \, ,
\label{1}
\end{eqnarray} 
$ \vec E$ - is the intensity of electric field and $\vec D$ and $\vec B$ 
are the electric and magnetic induction vectors,
t is the time , c is the speed of light. All the properties of the medium 
 are reflected
in the relation between $\vec B, \vec E$ and $\vec D$.
Eqs.(\ref{1}) would  suffice  to  describe  the   $\gamma$-beam 
propagation
in a medium and such a property as the intensity of magnetic field is not  
needed \cite{LL_ME, AG}.  
We represent the
relation between $\vec D$ and $\vec E$ in the form
\begin{equation}
 D_i(\omega)=\varepsilon_{ij}E_j(\omega)\, , \; \; \; (i,j = 1,2,3 )\, ,
\label{2}
\end{equation} 
where $ \varepsilon_{ij}= \varepsilon'_{ij} + i\varepsilon''_{ij}$
is the complex permittivity tensor and 
$\omega$  is the frequency of   $\gamma$-quanta.

 By way of example of the anisotropic medium let us consider
the superposition of the two linearly polarized laser waves, moving in the
same direction. The frequencies of these waves (photon energies) are 
different. 
  In order to determine the permittivity tensor in the case of a 
monochromatic field (high energy $\gamma$-beam)  
$ \vec E_0 e^{i(\vec k \vec r - \omega t)}$  propagating in above-mentioned 
laser medium , where 
 $ \vec k $ is the wave vector of the $\gamma$-quanta, we find the
average energy lost by the electromagnetic wave per unit volume $V$ and 
per unit time  \cite{LL_ME,AG}
\begin{equation}
\tilde q ={1\over {4\pi V}} \int_V \vec E {{\partial \vec D}\over {\partial t}} dV =
{{i\omega}\over {16\pi }} (\varepsilon_{ij}^\ast - 
\varepsilon_{ji})E_{j}^\ast E_{i} \, .
\label{3}
\end{equation}
The mechanism by which the wave loses energy is $e^+e^-$-pair production
in the field of the laser wave \cite{RN}.
The process is determined primarily by the transverse part of the permittivity
tensor, while the longitudinal components of the tensor are higher-order
infinitesimals in the interaction constant  $\alpha$ \cite{BKS,BT}. 
Taking this into account, and in the coordinate system one axis of which
is oriented parallel to the wave vector of $\gamma$-quanta, we have from
(\ref{3}) 
\begin{eqnarray}
\tilde q = {{i\omega J}\over {4}} \{({\varepsilon_{11}}^\ast - 
\varepsilon_{11})
(1+\xi_3) +({\varepsilon_{12}}^\ast - \varepsilon_{21})(\xi_1 - i\xi_2) +
  ({\varepsilon_{21}}^\ast- \varepsilon_{12})(\xi_1 +i\xi_2)+ \nonumber \\
  ({\varepsilon_{22}}^\ast -\varepsilon_{22})(1-\xi_3)\} \, ,
\label{4}
\end{eqnarray} 
where  $J = (E_{1}{E_{1}}^\ast +E_{2}{E_{2}}^\ast )/8\pi, \xi_i$ are Stokes
parameters of $\gamma$-beam. On the other hand, knowing the cross section 
 $\sigma_{\gamma \gamma}$ of the pair production process, we can write
\begin{equation}
\tilde q = 2cJ \{ n_{l,1} \sigma_{\gamma \gamma,1} 
+ n_{l,2} \sigma_{\gamma \gamma,2}\}
 \, , 
\label{5}
\end{equation}
where 
$n_{l,1}$  is the number  of photons per volume unit of the laser wave with the
linear polarization equals $P_{l,1}$, and $n_{l,2}$ is the similar
 number for second
wave with the linear polarization equals $ P_{l,2}$, 
$\sigma_{\gamma \gamma,1}$ and
$\sigma_{\gamma \gamma,2}$ are corresponding cross sections for $e^+e^-$ -pair
production in $\gamma \gamma$ -interactions, $P_{1,1},P_{3,1}$ and 
$ P_{1,2}, P_{3,2}$ are the Stokes parameters of the laser waves
($P_{1,1}^2 +P_{3,1}^2 = P_{l_1}^2,\,P_{1,2}^2+P_{3,2}^2 =P^2_{l_2}$).   
Factor 2 in this formula is due to the counter-motion of the $\gamma$-beam and
laser wave. Note, that the Eq.(\ref{5}) is true, when the intensity of
laser wave are not high (see below). 

We  can write the cross section of $e^+e^-$-pair production in the
following form  \cite{BKS,KS,B_DE,RN}  
\begin{eqnarray}
 \sigma_{\gamma \gamma}(z) = 
\sigma_0(z)+ \sigma_l(z)(\xi_1 P_1 +   \xi_3 P_3 )  \, , \, \; 0< z \le 1 
\,,
\label{6}\\
\sigma_0(z)= \pi r_e^2 z\{ (1+z-z^2/2) L_{\_} - \sqrt{1-z}(1+z)\} \, ,
\label{7}\\
\sigma_l(z)= {\pi r_e^2 z^3 \over 2} (L_{\_} + 2 \sqrt{1-z}/z) \, ,
\label{8} \\
L_{\_} =\ln {{1 + \sqrt{1-z}}\over \ {1-\sqrt{1-z}}} \, \nonumber
\end{eqnarray}
 where $z={{m^2c^4}\over {E_{\gamma} E_l}}$ is the invariant variable,
$E_{\gamma}$ is the $\gamma$-quantum energy, $m$ and $r_e$ are
 the mass and classical radius of electron,
$E_l$ and
$ P_1, P_3$  are the energy and Stokes parameters of the 
laser photon. It is well known that the pair production is a threshold
process and, because of this, the laser wave is a transparent medium for
$\gamma$-beam , when $E_{\gamma} E_l < m^2 c^4$ or
$z>1$. There are two different photon energies $E_{l,1}$ and $E_{l,2}$ in 
the case of 
dichromatic laser wave. Because of this, it is convenient to use the two
corresponding invariant  variable $z_1={{m^2c^4}\over {E_{\gamma} E_{l,1}}}$ 
and $z_2={{m^2c^4}\over {E_{\gamma} E_{l,2}}}$. It is evident that
$z_1/z_2 = E_{l,2}/E_{l,1}$.
Comparing Eqs.(\ref{4}) and  (\ref{5}) we can find the components
of permittivity tensor caused by $\gamma$-quanta absorption. 

Then we can determine the other components  of the tensor 
with the help of the following dispersion relations \cite{AG} 
\begin{equation}
\varepsilon'_{ij}- \delta_{ij}= {2\over \pi} 
{\cal P}\int_{0}^{\infty} {{x\varepsilon''_{ij}(x)\, dx}
\over {x^2-\omega^2}}
\label{9} \, ,
\end{equation}
\begin{equation}
 \varepsilon''_{ij}= -{{2\omega}\over {\pi}} {\cal P}\int_{0}^{\infty}
{{(\varepsilon'_{ij}-\delta_{ij}) \, dx}\over { x^2-\omega^2}}
\, ,
\label{10}
\end{equation}
where $\delta_{ij}$  is the Kronecker $\delta$-function. 
Comparing Eqs.(\ref{4}) and (\ref{5}), we get
\begin{eqnarray}
\varepsilon''_{11}+ \varepsilon''_{22}=4c (n_{l,1}\sigma_0(z_1) + n_{l,2} \sigma_0(z_2)) /\omega \, ,
\label{11} \\
\varepsilon''_{11}-\varepsilon''_{22} =4c (n_{l,1}\sigma_l(z_1) P_{3,1} + n_{l,2}\sigma_l(z_2) P_{3,2})  / \omega \, , 
\label{12}  \\
\varepsilon''_{12}+\varepsilon''_{21} =4c (n_{l,1}\sigma_l(z_1) P_{1,1} + n_{l,2}\sigma_l(z_2) P_{1,2})  / \omega \, ,       
\label{13} \\
\varepsilon'_{12}=\varepsilon'_{21}
\label{14}
\end{eqnarray}
It easy to verify that $\varepsilon_{12}=\varepsilon_{21}$.
The same result  is evident from the theory of generalized receptivity
 \cite{LL_SP}.

The other components  of the permittivity tensor  can be calculated 
with the help of  relations (\ref{9} )-(\ref{10}).
 The results of calculations of the components  $\varepsilon_{ij}$
 for the arbitrary coordinate system, one axis of which is oriented
parallel to the wave vector of the $\gamma$-quanta, are presented below
\begin{equation}
\varepsilon'_{11}-\varepsilon'_{22}=
{\alpha \over {2\pi E_o^2}}(<E^2_1> P_{3,1} z^2_1{F_1}^{\prime}(z_1)+ <E^2_2> P_{3,2} z^2_2{F_1}^{\prime}(z_2))
\label{15}
\end{equation}
\begin{equation}
\varepsilon'_{11}+\varepsilon'_{22} =2+ 
{2\alpha \over {\pi E_o^2}}
(<E_1^2>  z^2_1{F_2}^{\prime}(z_1, 1) + <E_2^2> z^2_2{F_2}^{\prime}(z_2, 1)) \, ,
\label{16}
\end{equation}
\begin{equation}
\varepsilon'_{12}=\varepsilon'_{21}=
{\alpha \over {4\pi E_o^2 }}(<E^2_1> P_{1,1} z^2_1{F_1}^{\prime}(z_1)+ <E^2_2>  P_{1,2} z^2_2{F_1}^{\prime}(z_2))
\label{17}
\end{equation} 
\begin{equation}
\varepsilon''_{11}-\varepsilon''_{22} =
-{\alpha \over 4 E_o^2  }(<E^2_1>  P_{3,1} {F_1}^{\prime \prime}(z_1)+ 
<E^2_2>  P_{3,2} {F_1}^{\prime \prime}(z_2)) \, ,
\label{18}
\end{equation}
\begin{equation}
\varepsilon''_{11}+\varepsilon''_{22}=
{\alpha \over E_o^2 }( <E^2_1> {F_2}^{\prime \prime}(z_1, 1)  +
 <E^2_2> {F_2}^{\prime \prime}(z_2, 1 ) \, ,
\label{19}
\end{equation}
\begin{equation}
\varepsilon''_{12} = \varepsilon''_{21} =
-{\alpha \over 8 E_o^2 }(<E^2_1> P_{1,1} {F_1}^{\prime \prime}(z_1)+ 
<E^2_2> P_{1,2} {F_1}^{\prime \prime}(z_2)) \, ,
\label{20}
\end{equation}
where $<E^2_i>= 4\pi n_{l,i} E_{l,i} \, (i=1,2) $ is the mean square 
of intensity electric field for every laser wave  and the functions  
    $F'_1, F'_2,  F''_1, F''_2, $ are equal to:
\begin{equation} 
{F_1}^{\prime }(z) =
\cases{ 
[\sqrt{1-z}+{z\over 2}L_-]^2+[\sqrt{1+z}-{z\over 2}L_+]^2 -
{{\pi^2z^2}\over {4}},\; 0<z \le 1,  \cr 
 -[\sqrt{z-1} -z \arcctg\sqrt{z-1}]^2+[\sqrt{1+z}- {z\over 2}L_+]^2 
, \; z>1. \cr}
\label{21}
\end{equation}
\begin{equation}
{F_2}^{\prime }(z,\mu ) =
\cases{ 
-2-\mu -(1+\mu (z-{z^2\over 2})){1\over 4} L^2_- 
-(1-\mu (z+{z^2\over 2})){1\over 4}L^2_+ + \cr
+{{(1+\mu z )\sqrt{1-z}}\over{2}}L_- -{{(\mu z-1)\sqrt{z+1}}\over {2}}L_+ 
+ {\pi^2 \over 4}(1+\mu (z-{{z^2}\over 2})),\; 0<z \le 1, \cr
-2-\mu  +(1+\mu (z-{z^2\over 2})) \arcctg^2(\sqrt{z-1}) 
-(1-\mu (z+{z^2 \over 2})){1\over 4}
L^2_+ + \cr
+ (1+\mu z)\sqrt{z-1} \arcctg\sqrt{z-1} -{{(\mu z-1)\sqrt{1+z}}\over{2}} L_+ 
,\; z>1.  \cr }
\label{22}
\end{equation}
\begin{equation}
{F_1}^{\prime \prime}(z)=
\cases{ 
z^4 (L_- + {{2\sqrt{1-z}}\over{ z}}) , \; 0< z \le 1,  \cr
0, \; z>1. \cr}
\label{24}
\end{equation}
\begin{equation}
{F_2}^{\prime \prime}(z,\mu )=
\cases{ 
z^2((1+\mu (z-{z^2\over 2}))L_- -\sqrt{1-z}(1+\mu z)), \; 0<z \le 1,  \cr
0, \; z> 1 \cr}
\label{25}
\end{equation}
 The function   $L_+$ is equal to:
\begin{equation}
\nonumber L_+ =ln{{\sqrt{1+z}+1}\over {\sqrt{1+z}-1}} \, .
\end{equation}
 The constant  $E_o = {{m^2c^3}\over {e\hbar}}$ 
is the critical field of quantum electrodynamics.
The presented here data completely determine the permittivity tensor
for high energy $\gamma$-quanta traversing a dichromatic
linearly polarized  bunch of laser photons.

In a number of problems in  optics it is more convenient to
employ the tensor    $\eta_{ij}$ , which is inverse of the tensor
$\varepsilon_{ij}$. When $|\varepsilon_{ij} - \delta_{ij}| \ll 1$, 
these tensors are related by
\begin{equation}
\eta_{ij}+\varepsilon_{ij} = 2\delta_{ij}.
\label{26}
\end{equation}
 
 It should be noted the following peculiarities of the permittivity tensor:  

1) Our description can be extended   
to the case when the laser bunch  is superposition more then two linearly   
polarized waves.  It is obvious that the analogous terms should be added
in the tensor components in these cases.

2) Let us that $E_{l,1}> E_{l,2}$. Then the laser bunch is a transparent 
medium at $z_1 >1$. In this case the all components $\varepsilon''_{ij}$ are
equal to zero.

3) In the general case the symmetric complex tensor $\varepsilon''_{ij}$
does not reduce to principal axes \cite{LL_ME,AG,MMF} (i.e., there does
not exist a coordinate system in which the tensors $\varepsilon'_{ij}$ 
and $\varepsilon''_{ij}$ are simultaneously diagonal).

4)In the case, when the two waves have the same direction of linear
polarization or their polarizations are orthogonal, the tensor
$\varepsilon_{ij}$ can be reduced to  a diagonal form.
The permittivity tensor for monochromatic linearly polarized wave 
can be reduced to the diagonal form for the all time. 

Everything said above is also true for tensor $\eta_{ij}$.

\section{ Refractive indices of $\gamma$-quanta  }
The main problem of optic of anisotropic  medium is to
investigate the propagation of monochromatic plane waves, characterized
by definite values of the frequency $\omega$ and wave vector $\vec k$.
Such waves, satisfying a homogeneous wave equation, are called normal
electromagnetic waves \cite{AG}, and they have the form   
$$\vec E = \vec E_0e^{
i(\vec k \vec r - \omega t)}, \vec k = \omega \tilde n \vec s/c \, , $$
where  $\vec E_0$ is the complex vector, independent of coordinates 
$\vec r$  and time, $\tilde n$ is the complex index of refraction
and  $\vec s =\vec k/|k|$ is a real unit vector. The vectors  $\vec D$ 
 and $\vec B$ have the same form.
  
 From Maxwell's equations (1) we obtain the wave equation
$$ rot rot \vec E + {1\over c} {{\partial^2 \vec D} \over {\partial t^2}} =0\,.  
$$
 Taking into account the relation between $\vec D$ and $\vec E$ in a 
system of coordinates in which the axis $x$ is oriented parallel to the
wave vector, we obtain
\begin{eqnarray}
\nonumber
\eta_{11} {{\partial^2 D_1}\over {\partial x^2}} + \eta_{12} 
{{\partial^2 D_2}\over {\partial x^2}} - {1\over c^2}
{{\partial^2 D_1}\over {\partial t^2}} =0, \\
\eta_{21} {{\partial^2 D_1}\over {\partial x^2}} +\eta_{22}
{{\partial^2 D_2} \over {\partial x^2}} -{1\over c^2}
{{\partial^2 D_2}\over {\partial t^2}}=0 .
\label{27}
\end{eqnarray} 
For a monochromatic plane wave it follows from these equations that
\begin{equation}
(\tilde n^{-2} \delta_{ij} - \eta_{ij})D_j=0,\; i,j=1,2.
\label{28}
\end{equation}
From the condition that the two homogeneous equations are compatible, we
find the index of refraction of the  $\gamma$-quanta
\begin{equation}
\tilde n^{-2} = {S\over 2} \pm \sqrt{{S^2 \over 4}- D_{\eta}}
=(\eta_{11}+\eta_{22})/2 \pm 
\sqrt{(\eta_{11}-\eta_{22})^2/4 + \eta_{12}\eta_{21} }
\, ,
\label{29}
\end{equation}
where $S$ and $D_{\eta}$ are, respectively, the trace and  determinant
of the matrix $\eta_{ij}$.  Thus, in the general case the $\gamma$-beam  
propagates through the
laser wave as the superposition of two electromagnetic waves with different
refractive indices. Note, that the two roots of (\ref{30}) which have
form $-1 + small \; quantity $, are superfluous. They  correspond to
the $\gamma$-quanta motion in the reverse direction.     
  
In the general case the refractive indices are  complex values.
However, they are real values, when  the laser bunch is  a transparent
medium for  $\gamma$-quantum (all components of the permittivity tensor are
real number in this case). 
Fig.1 illustrates the refractive indices as functions
of the invariant variable~$z$ (the laser wave parameters are in the caption).

\section{ Polarization properties of $\gamma$-beam propagation in laser wave}
Here we consider the polarization properties one of two normal 
electromagnetic waves. 
From  dispersion equations (\ref{29}) we find the ratios of the components
of the vector $\vec D$
\begin{equation}
{D_1\over D_2} =\kappa = {{\tilde n^{-2} - \eta_{22}}\over {\eta_{21}}} =
{{|D_1|}\over {|D_2|}}e^{i\delta},
\label{30}
\end{equation} 
where $\delta$ is the phase shift between $D_1$ and $D_2$. This ratio
$\kappa$ can be reduced to zero or to the form $\kappa = i\rho$
(since $|D_1| |D_2| \sin \delta =b_1 b_2$, where $b_1$ and $b_2$ are the
semiaxes of the ellipse and $|\rho| = b_1 / b_2$ \cite{LL_TF})  by the
rotation of the coordinate system around  the wave vector of $\gamma$-quanta
(the  $x$-axis is constantly aligned with the wave vector). The first case 
corresponds to the propagation of a linearly polarized wave and the
second case corresponds to an elliptically polarized wave; in addition,
$\rho > 0 (\rho <0)$ corresponds to left (right) - hand polarization of
$\gamma$-quanta. 

  The different cases of $\gamma$-beam propagation in anisotropic medium, 
which optical properties described by the symmetric tensor, were
considered in paper \cite{MMF}. In the case when permittivity tensor
may be reduce to principal axes (i.e., there is a coordinate system
in which $\varepsilon_{12}=0$) the normal electromagnetic waves are
linearly polarized. In general case the permittivity tensor does not
reduce to principal axes and the normal waves are elliptically polarized.
The propagation of $\gamma$-beam, which is a superposition of the two
linearly polarized waves, was considered in a number papers \cite{KS,MV,MMF1}.
Because of this, in the following we will consider the case when 
$\gamma$-beam is the superposition of the two elliptically 
polarized normal waves.

 In the case under consideration the 
refractive indices are the  complex values. Because of this, the value
$\kappa$ is also complex and we get the following relation between 
two normal waves
\begin{equation}
 \kappa^{(1)} \kappa^{(2)} = - 1 \, ,
\label{31}
\end{equation}   
where the indices in parentheses refer to the waves with refractive indices
$\tilde n_1$ and $\tilde n_2$. In what follows we will use only one  
of two values, namely, the $\kappa= \kappa^{(1)}$ (without pointing any
indices). 
 In our case one can obtain
\begin{eqnarray}
 D_1^{(1)}D_1^{(2)} + D_2^{(1)}D_2^{(2)} = 0\, ,
\label{32} \\
 D_1^{(1)}D_1^{*(2)}+D_2^{(1)}D_2^{*(2)}= D_2^{(1)}D_2^{*(2)}(\kappa^{*} - 
\kappa )/\kappa^{*} \, ,
\label{33}
\end{eqnarray}
where the indices in parentheses refer to waves with refractive indices
$\tilde n_1$ and $\tilde n_2$. From here we can see that $\vec D^{(1)}$  
and $\vec D^{(2)}$ vectors are  orthogonal but $\vec D^{(1)}$ and 
$\vec D^{*(2)}$ vectors  are not orthogonal if the value 
$\kappa^{*}-\kappa $ is not equal to zero.     
     Let us name the Stokes parameters of the normal wave with the
refractive indices $\tilde n_1$ and $\tilde n_2$ respectively  as 
$X_1, X_2, X_3$ and $Y_1,Y_2,Y_3$. Then we get
\begin{eqnarray}
X_1={\kappa + \kappa^{*} \over 1+\kappa \kappa^{*}} \, , 
\label{34} \\
X_2= {i(\kappa - \kappa^{*}) \over 1+ \kappa \kappa^{*}} \, , 
\label{35} \\
X_3= {\kappa \kappa^{*} -1 \over 1 +\kappa \kappa^{*}} \, .
\label{36}
\end{eqnarray}
We have also the following relations $Y_1=-X_1, Y_2=X_2, Y_3= -X_3$. 
The angle of ellipse turn  $\varphi$ is found from relation 
$tg2\varphi =X_1/X_3 = Y_1/Y_3$.

Fig.2 illustrates the results of calculations of $|P_{circ}|=|X_2| $ 
  as functions of the invariant variable $z_1$ under various conditions.

\section{ $\gamma$-quanta propagation in the laser wave}
 Now we can find the relations describing the variations of intensity
and Stokes parameters of $\gamma$-quanta propagating in the
uniform ($n_l =const$) laser wave. 
Then representing the $\gamma$-beam as the superposition of two
normal waves corresponding to the polarization state of a laser wave
we get the following relations 
\begin{eqnarray} 
J_{\gamma}(x)= J_1(x)+J_2(x)+2J_3(x) \, , 
\label{37} \\
\xi_1(x)= (X_1 J_1(x) + Y_1 J_2(x)+ p_1 J_4(x))/J_{\gamma}(x)\, ,
\label{38} \\
\xi_2(x) =(X_2 J_1(x) + Y_2 J_2(x) + p_2 J_3(x))/J_{\gamma}(x) \, , 
\label{39} \\
\xi_3(x)=(X_3 J_1(x) + Y_3 J_2(x) +p_3 J_4(x))/J_{\gamma}(x) 
\, ,\label{40}
\end{eqnarray}
where  $J_{\gamma}, \xi_1, \xi_2, \xi_3$  are the intensity and 
Stokes parameters of $\gamma$-quanta on
the laser bunch thickness equal to x.  The partial intensities $J_i(x),
(i=1-4) $ have the following form ( the physical sense of these values is
easy to understand, if the relation  
$(\vec{D}^{(1)} + \vec{D}^{(2)})(\vec{D}^{*(1)}+\vec{D}^{*(2)})$ is written
in the component-wise form)
\begin{eqnarray}
J_1(x)=J_1(0)\exp(-2\Im(\tilde n_1)\omega x/c) \, ,
\qquad \qquad \qquad \label{41} \\
J_2(x)=J_2(0)\exp(-2\Im(\tilde n_2)\omega x/c) \, ,
\qquad \qquad \qquad \label{42} \\
J_3(x)=\exp(-\Im(\tilde n_1 +\tilde n_2 ) \omega x/c)\{J_3(0)\cos(\Re(\tilde n_1 -
\tilde n_2)\omega x/c) + \\ 
+  J_4(0) \sin(\Re(\tilde n_1 - \tilde n_2) \omega x/c)\} \qquad \qquad \qquad
 \,,\label{43} \nonumber \\
J_4(x)=-\exp(-\Im(\tilde n_1+ \tilde n_2) \omega x/c)\{J_3(0) \sin(\Re(\tilde n_1 -
\tilde n_2) \omega x/c) - \\
-J_4(0)\cos(\Re(\tilde n_1- \tilde n_2)\omega x/c)\} \qquad \qquad \qquad 
\,. \label{44} \nonumber
\end{eqnarray} 
The initial partial intensities are defined from the following relations
\begin{eqnarray}
J_1(0)={\xi_2(0)-q \over 2(X_2-q)} +
{\xi_3(0)-f\xi_1(0) \over 2(X_3-fX_1)} \, ,\label{45} \\
J_2(0)={\xi_2(0)-q \over 2(X_2-q)} -
{\xi_3(0)-f\xi_1(0) \over 2(X_3-fX_1)} \, ,\label{46} \\
J_3(0)=1/2 -  {(\xi_2(0) - q) \over 2(X_2-q)} \,, \label{47}\\
J_4(0)={\xi_1(0)X_3- \xi_3(0)X_1 \over p_1X_3 - p_3X_1} \,.
\label{48}
\end{eqnarray}
The relations between $X_i$ and $Y_i$ values were used, because of this 
the $Y_i$-values are absent in  Eqs.(\ref{45})-(\ref{48}). Besides, we
assume that $J_{\gamma}(0)=1$. The parameters $f,q,p_1,p_2,p_3$ have
the following form:
\begin{eqnarray}
p_1= {2i(1-\kappa \kappa^{*}) \over \kappa^{*} -\kappa}\,, \quad 
p_2= {2i(1+\kappa \kappa^{*}) \over \kappa^{*} -\kappa}\,, \quad
p_3= {2i(\kappa+\kappa^{*}) \over \kappa^{*} -\kappa} \,. \nonumber \\
f=p_3/p_1 \, , \qquad q=p_2/2\, . \nonumber 
\end{eqnarray}

Eqs.(\ref{37})-(\ref{40})  describe the general case of
$\gamma$-beam propagation, when the variations intensity and
Stokes parameters are determined by the imaginary values of
refractive indices and the difference of their real quantities. 
However, these equations do not described such cases, when  
the relation $\kappa + \kappa^{*} =0$ takes  place. In the case
$\kappa + \kappa^{*} =0$, one can use the known relations from papers
\cite {KS,MV,MMF1} or one can find limits of the obtained here 
Eqs.(\ref{37})-(\ref{40}). For example, one can offer 
$\kappa= \delta + i\rho$ and $\delta$ direct to zero. 

Figs.3,4 illustrate the variations of Stokes parameters and intensity     
of an initially unpolarized $\gamma$-beam moving in the dichromatic laser wave.

\section{ Influence of the laser wave intensity on $\gamma$-quanta
propagation }
The influence of the laser wave intensity on the  $e^+e^-$-pair     
production was studied in a number of papers (see Ref.\cite{BKS} and
literature therein).
 The degree of intensity of a dichromatic laser wave can be characterized  
by the dimensionless parameter \cite{BKS}
$\xi^2=\xi^2_1 +\xi^2_2$ where
 $\xi^2_1={<E^2_1> \over {E_o}^2} {m^2c^4 \over {E_{l,1}}^2}$ and
$\xi^2_2={<E^2_2> \over {E_o}^2} {m^2c^4 \over {E_{l,2}}^2}$.
Here we have considered the case of the relatively not high intensity of a
laser wave, when $\xi^2 \ll 1$. The results of  papers \cite{BKS,RN} 
are allowed one to write down the components of permittivity tensor taking into
account the first terms of the  expansion of the intensity in 
Taylor series.  Thus we  have found that the already obtained  components of
the tensor should be transformed with the help of the following
simple rules. Firstly, the variables $z_i, i=1,2$ are substituted by the 
variables $\tilde z_i =z_i(1+\xi^2)$. Secondly, the value $E_o$ (the critical
field)  in Eqs.(\ref{15})-(\ref{20}) is substituted by the
value $\tilde E_o ={m^2c^3(1+\xi^2) \over e \hbar}$. 
Thirdly, the functions  
 $F'_2(z_i,1), F''_2(z_i,1), F'_1(z_i), F''_(z_i) $ 
are substituted by the functions 
 $F'_2(\tilde z_i,\mu), F''_2(\tilde z_i, \mu),F'_1(\tilde z_i), 
 F''_1(\tilde z_i),$
where  $\mu =1 / (1+\xi^2)$. The new condition for the pair
production threshold follows from these rules. It is $\tilde z_i < 1 $,
where i is index of wave with the most photon energy.  
It means that the threshold energy of $\gamma$-beam  enhances
(at the fixed frequency of laser photons). In the strict sense
the field of  application  of these more refined relations 
satisfies the condition $\xi^2 \ll 1$.   Nevertheless, we can receive
the important information in this case \cite{MV}. 

In the case, when the parameter $\xi^2 \gg 1$, the pair production process
is similar to an analogous process in the constant electromagnetic field.
The permittivity tensor for these fields was found in Ref.{\cite{BKF}}
and some particular calculations of $\gamma$-quanta propagation
are in  \cite{MMF2}. 

\section{Discussion}       
  The optical properties of an anisotropic medium can be described by the use 
of the symmetric permittivity tensor. In general case the tensor components
are complex values. It means, broadly speaking, that the permittivity
tensor does not reduce to principal axes with the result that 
the normal electromagnetic waves (the eigenfunctions of the problem)
present two elliptically polarized  waves.  Because  of  this  some 
peculiarities
in $\gamma$-quanta propagation in anisotropic medium are appeared.
  The simplest example of such a medium of the general type is the
dichromatic laser wave involving two linearly polarized waves with
different frequencies.  Other example is a single crystal oriented
in region of the coherent pair production process \cite{BKS}. 
The permittivity tensor and polarization characteristics of normal waves
in single crystals were obtained in paper \cite{MMF}, and it was
shown that  orientation regions are in single crystals, where the
circular polarization of normal waves is high ($\approx 90\%$ in maximum).   
However in single crystals the components of permittivity tensor
are a sum of sufficiently large number of terms that make the 
investigation more difficult. Note that Eqs.(\ref{37}-\ref{40}) 
for variations of $\gamma$-beam intensity and Stokes parameters are
true for an arbitrary anisotropic medium, when $\kappa-\kappa^* \ne 0$
 
We made some calculations of $\gamma$-beam propagation in the field of
dichromatic wave in the case, when the frequency ratio $E_{l,2}/E_{l,1}$
is equal to 2 (see Figs.(1-4)). We take the values $P_{1,1}=1, P_{3,2}=1$
for the polarization state of the dichromatic wave. It means that
the angle between  directions of these two polarizations is equal to 45
degree. One can see from Fig.2 that value
$|P_{circ}|= |X_2|$ depends on ratio of the electric field intensities 
 $r= \sqrt{<E_2^2>/<E_1^2>}$ . The value $|X_2|=1$, when r=0.658 or 4.12
(curves 1 and 4 on Fig.2). The refractive indices for r=0.658 are shown
on Fig.1. One can see that the real and imaginary parts of refractive
indices of two normal electromagnetic waves are equal in magnitude
at $z_1=1.43$. This  is the so-called in classical crystal optics
 case of  singular axis \cite{LL_ME,AG}.
  
 Now we can make a conclusion that in single crystals the high degree
of circular polarization of normal waves \cite{MMF} is due to 
the common action of the two "strong" crystallographic  planes with  the 
45 degree angle between them. The (110) and (010) planes are responsible
for the effect at the conditions of the cited paper.   
 
 Fig.3 illustrates the variations of Stokes parameters as functions
of the laser bunch thickness. The behavior of these curves it is easily
to understand. As already noted, the $\gamma$-beam in a medium can be
presented as a superposition of two normal electromagnetic waves with
the different refractive indices. Because of this, one normal wave is
adsorbed more then another wave and after propagation of some thickness $x$
only this wave would then be left behind and $\xi_1(x) \approx X_1,
\xi_2(x) \approx X_2 , \xi_3(x) \approx X_3$  or $\xi_1(x) \approx Y_1,  
\xi_2(x) \approx Y_2 , \xi_3(x) \approx Y_3$. Referring to Fig.3 it will
be observed that such a thickness is enough large when $ 1< z_1 <1.45$. 
The reason is that the difference of imaginary parts of refractive
indices is a small value in the pointed region of the variable $z_1$(see Fig.1).

The dichromatic wave or single crystals \cite{MMF} are sensitive to the
circular polarization of $\gamma$-beam (see Figs.3,4). So, initially an
unpolarized $\gamma$-beam moving in an anisotropic medium become 
circularly polarized one. This case is differ from the known case \cite{C}
of the $\gamma$-beam propagation in the anisotropic medium. As is shown in
 Ref.{\cite{C}} the only linearly polarized beam is transformed in
circularly polarized one. In principle, the single crystal sensitivity to
a circular polarization can be used for determination of polarization of
high energy ( in tens GeV and more) $\gamma$-quanta and electrons.

It is believed that a $\gamma$-beam propagation in the linearly polarized
monochromatic laser
wave moving in the  magnetic field (normally to it direction)
is similar to  such a propagation in a dichromatic wave. The another
analogous example is two monochromatic laser waves with  the  equal 
frequencies   
moving at an nonzero angle between them.  
On the other hand we would expect that high energy
$\gamma$-quanta, propagating over cosmological distances, can be polarized
due to their interaction with magnetic fields and/or extragalactic starlight
photons \cite{SS}. With this point of view one can explain the anisotropy
in electromagnetic propagation over cosmological distances \cite{NR_A}, 
as might  appear  at  first  sight.  However  this  explanation  is 
required
a high electromagnetic energy density in the cosmic space ($>0.1 \, erg\, c
m^{-3}$, if take the  birefringence scale of order $10^{27}$ cm \cite{NR_A}).
This estimate  follows from Ref.\cite{MV} where asymptotic values of 
the refractive indices are calculated.

The propagation of $\gamma$-quanta through a laser wave (when $\xi^2 \ll 1$)
is similar to the same process in single crystals for the region of 
coherent pair production. For example, the permittivity tensor in
single crystals \cite{MMF} is  determined by the functions 
 $F'_1, F'_2, F''_1, F''_2 $  as in a laser wave. However,
the existence of some frequencies of pseudophotons and incoherent pair
production in single crystals is the main difference between these two  
cases. Note that the permittivity tensor components (see also Ref.\cite{MV}) 
can be presented as a linear combinations
of the invariant helicity amplitudes for the forward light by light
scattering \cite{KS,B_DE,LL_KE}.

Note  that  there  are  no  experiments  yet  in  support  of  the 
transformation
of $\gamma$-beam polarization  in single crystals and laser waves. Nevertheless, 
a number of proposals on the investigation and utilization of this
phenomenon \cite{BT} is available. 

The author would like to thank S.Darbinian for usefull discussion.

\vspace{3.2 cm}

\newpage

\begin{figure}[h]
\begin{center}
\parbox[c]{13.5cm}{\epsfig{file=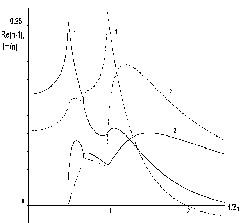,height=10cm}}
\parbox[c]{15cm}{\caption{
 The real (1)  and imaginary (2) parts of  refractive indices for dichromatic 
laser wave ($P_{1,1}=1$, $P_{3,2}=1$, $r=0.658$, $z_2/z_1=2$)
as functions of the invariant parameter $z_1$. For obtaining absolute quantity
the ordinate value is myltiplied by the factor $E_{o}^{2} / \alpha <E^{2}_{1}>$.
              }}  
\end{center} 
\end{figure}

\begin{figure}[h]
\begin{center}
\parbox[c]{13.5cm}{\epsfig{file=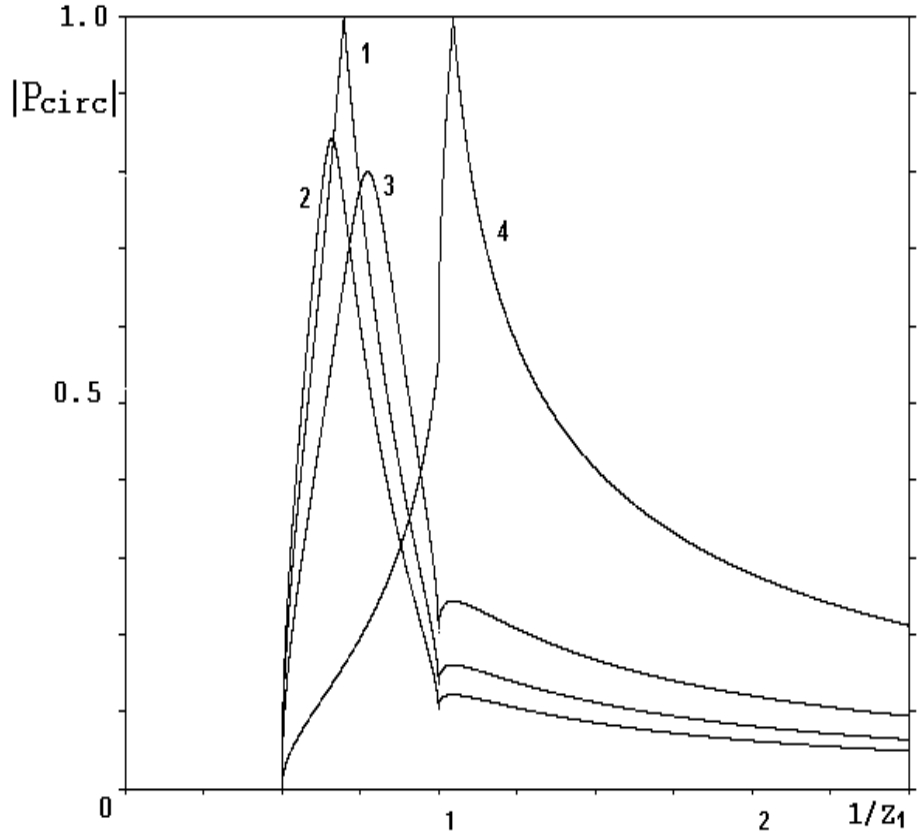,height=12cm}}
\parbox[c]{15cm}{\caption{
 The absolute value of normal waves circular polarizition
$|P_{circ}|= |X_2| $  as function of the invariant parameter $z_1$.   
The polarization state of laser wave is as in Fig1.
The ratios $r$ are  equal to 0.658,0.5,1.0 and 4.12 for
curves 1-4. The ratio $z_1/z_2$ is equal to 2.
              }}  
\end{center} 
\end{figure}

\begin{figure}[h]
\begin{center}
\parbox[c]{13.5cm}{\epsfig{file=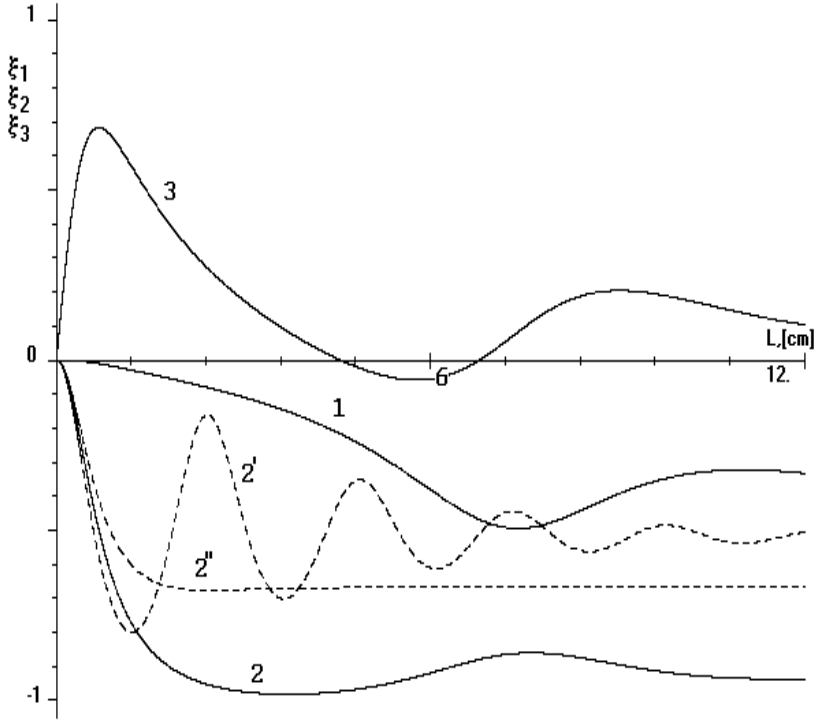,height=10cm}}
\parbox[c]{15cm}{\caption{
 The Stokes parameters of initialy unpolarized $\gamma$-beam as
functions of a laser bunch thickness. The polarization state of laser
wave is as in Fig.1. The number near each  curve corresponds to number i of
Stokes parameter $\xi_i$. The solid curves correspond to $z_1=1.4$.
The curves $2'$ and $2"$ correspond to $z_1= 1.2,\, 1.6$.   
$r=0.658$, $z_1/z_2=2$, $n_{l,1}=5\,10^{25}cm^{-3}$, $E_{l,1}=1eV$.
 }}  
\end{center} 
\end{figure}

\begin{figure}[h]
\begin{center}
\parbox[c]{13.5cm}{\epsfig{file=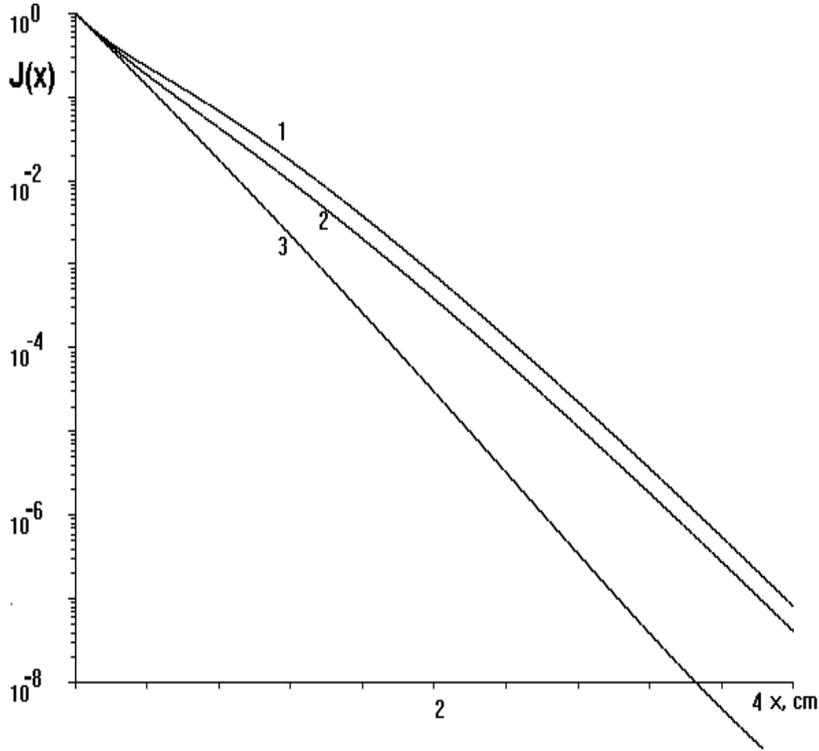,height=10cm}}
\parbox[c]{15cm}{\caption{
 The intensity of $\gamma$-beam as function of the laser bunch  
thickness. The polarization state of laser wave is as in Fig.1.
Curves 1-3 correspond to initial circular polarization $\xi_2(0)$ of 
$\gamma$-beam equals  1, 0 and -1. 
$r=0.658$, $z_1/z_2=2$, $n_{l,1}=5\,10^{25}cm^{-3}$, 
$E_{l,1}=1eV$. }}  
\end{center} 
\end{figure}

\end{document}